# Dislocation in Motion as the Dynamic Distribution of Elastic Field Singularity

A. Dutta[1], M. Bhattacharya[2], P. Mukherjee[2], N. Gayathri[2], G. C. Das[1], and P. Barat[2]

[1]Department of Metallurgical and Materials Engineering, Jadavpur University, Kolkata 700 032, India
[2]Variable Energy Cyclotron Centre, 1/AF, Bidhannagar, Kolkata 700 064, India

Plastic deformation of crystals is a physical phenomenon, which has immensely driven the development of human civilisation since the onset of the Chalcolithic period. This process is primarily governed by the motion of line defects, called dislocations. Each dislocation traps a quantum of plastic deformation expressible in terms of its Burgers vector[1,2]. Theorising the mechanisms of dislocation motion at the atomistic scales of length and time remains a challenging task on account of the extreme complexities associated with the dynamics. We present a new concept of modelling a moving dislocation as the dynamic distribution of the elastic field singularity within the span of the Burgers vector. Surprisingly, numerical implementation of this model for the periodic expansion-shrinkage cycle of the singularity is found to exhibit an energetics, which resembles that of a dislocation moving in the presence of the Peierls barrier[1-4]. The singularity distribution is shown to be the natural consequence under the external shear stress. Moreover, in contrast to the conventional assumption, here the calculations reveal a significant contribution of the linear elastic region surrounding the core towards the potential barrier.

Dislocations were initially studied by Vito Volterra[5] as purely mathematical problems within the framework of continuum mechanics long before they were proposed to be responsible for the plastic deformation of crystalline solids[6-8]. In his theory, a dislocation was defined mathematically as an elastic strain field, which is essentially singular over a locus of points known as the dislocation line. Later on, the celebrated Peierls-Nabarro (PN) model[3,4] witnessed phenomenal success as it indicated the origin of the potential barrier against the motion of dislocations emerging out of the discreteness of the crystalline structure. It employs a different formulation of the elastic field taking care of the nonlinearity in the dislocation core region and considers a distribution of infinitesimal dislocations over the entire plane of the lattice misfit. The model is further reinforced with the idea of the generalized stacking fault[9], variational formulation[10] and other important refinements[11-13]. Even though the solutions for the elastic fields from these two models converge in the far field region, they are widely different near the dislocation core[1]. In addition, the PN model removes the singularity of the elastic field at the origin[1]. Nevertheless, it must be noted that the concept of singularity in Volterra's dislocation is entirely mathematical and does not create any physical ambiguity as the crystal possesses a discrete atomistic construction and the dislocation line can be placed without passing through any atom. This dislocation line, defined in terms of the elastic field singularity, can be considered as a tag to specify the exact location of the dislocation in the solid. However, in crystals with discrete and periodic assembly of atoms, the idea of specifying the position of a dislocation bears a sense only at its static sites separated by the length of the Burgers vector. The exact position of the dislocation is inexpressible when the dislocation is moving from an equilibrium site to the next one, as the motion is continuous instead of hopping. Thus, it is justifiable to consider a moving dislocation as delocalized and spread out while it is present in between two adjacent static sites (Fig 1a). This suggests that the singularity of its strain field may also be distributed simultaneously so that this delocalization can directly influence the strain field in the solid. The robustness of this approach lies in the fact that it uses the well behaved solutions of the undistributed singularity, in conjunction with the principle of superposition. In the case of dislocation motion, the line singularity of the associated strain field can be distributed dynamically in a manner such that it becomes a line singularity again after a specific interval of time, when it has shifted from the previous line position by the Burgers vector. After having proposed this model of finite singularity distribution, we now face some essential questions. What is the energetics of this distribution process? Is it physically feasible? Is this mathematical concept capable of reproducing the mechanism of the real dislocation motion? In order to seek the answers, we implement this model numerically with simple assumptions and investigate the validity of the fundamental principles involved herein.

We perform the computation assuming the simple case of isotropic elasticity and a straight edge dislocation in infinite solid. We select a finite rectangular domain $\Omega$ with the dislocation at its centre as shown in Fig. 1b. This domain is divided into small square subdomains with the length of the Burgers vector (b) as their sides. We assume a simple homogeneous distribution of the singularity of the strain field of the edge dislocation over the span of one Burgers vector divided into 50 equal divisions (Fig. 1c). Consequently, as the dislocation proceeds by one Burgers vector, 99 discrete steps of singularity distribution are required. At each step of the distribution, the magnitude of the Burgers vector per unit length is normalized so that the total Burgers vector is conserved. The subdomains within a radius $r_c = 2b$ from the dislocation line are excluded from the calculation to avoid the singularity at the origin and the non-linearity of the core region. At any particular step of the singularity distribution, the strain in a subdomain is evaluated by superposing the strains contributed by each of the 50 discrete divisions of the Burgers vector. The elastic energy of each subdomain is calculated using the expressions for the strain field of an edge dislocation[1] and added up to obtain the total strain energy of the whole domain. The result of computation for a domain of dimensions (400b × 50b) is given in Fig. 1d with the Young's modulus E=1 (in arbitrary units) and the Poisson's ratio ν=0.33. The minimum elastic energy is at the line singularity (first step in Fig. 1c) and it ascends with the proceedings steps of the distribution. At the maximum extent of the distribution ($50^{th}$ step), when the singularity covers the entire span of one Burgers vector, the elastic energy is also at its peak and falls symmetrically over the remaining half cycle at the end of which the line singularity is shifted by one lattice spacing ($99^{th}$ step). The energy profile of Fig. 1d is instantly recognized as the potential barrier to dislocation motion, which is well known to impose a lattice resistance against the gliding dislocation. At this stage, it can be pointed out that the fundamental concept followed here is entirely different from the treatment in the PN model, where the misfit energy was essentially summed over the atom rows to exhibit the effect of the lattice discreteness. Instead, mere cyclic distribution of the strain field singularity is capable of demonstrating the existence of the potential barrier, which the dislocation must overcome to shift to the adjacent energy valley one Burgers vector away. However, a dislocation can move only under the influence of an applied shear stress. Therefore, the energetics of the system evolving out of the proposed model in the presence of applied shear stress must be compatible with the real physical phenomenon. We verify this by repeating the abovementioned computation procedure with applied shear stresses. In contrast to the previous case of zero external stress (Fig. 1d), the minimum of the elastic energy now shifts towards the state of more expanded distribution (Fig. 2) under external shear stress, implying that the configuration with the expanded singularity is the preferred and energetically stable. These results establish that the distribution of the elastic field singularity is a spontaneous process, whenever the system is subjected to applied shear stress. At a critical value of the stress ($\tau_{xy}$=0.013 in this case), the energy valley reaches to the next lattice site and this critical value is the counterpart of the well known Peierls stress in the PN model[3,4]. On increasing the shear stress further, the material fails and starts deforming plastically, thereby lowering the internal stress instantaneously. Consequently, the state of full expansion of the singularity becomes unstable and the singularity starts shrinking until it recovers back the state of the line singularity by the time it shifts by one Burgers vector with respect to the initial position. Interestingly, applying the components of stress other than $\tau_{xy}$ are not found to alter the potential barrier. On reversing the direction of $\tau_{xy}$ results in large potential barrier without showing any intermediate energy minimum. This is again appropriate in physical terms as it corresponds to a dislocation moving against the direction of applied stress, which is highly improbable.

Defining a moving dislocation as the dynamically delocalizing singularity of the associated elastic field is a fruitful way of understanding the mechanism of dislocation dynamics at the atomistic scale. It is projected as a conceptual tool independent of the existing conventional models. Although, the numerical implementation has been done only for the linear elastic region with the assumption of isotropy and a simple form of singularity distribution, the present work motivates further studies on this model to make it more quantitative in nature.

Figure Captions:

Fig 1: a. Schematic of a moving dislocation as delocalized in between two static sites. b. Schematic representation of the computational method. $\Omega$ is the rectangular domain with an edge dislocation at the origin O. A typical square subdomain is shown with the position vector **r** from the origin to the center of the subdomain. Calculations are done using 200 such subdomains along the x-direction on both sides of O. The number of subdomains along the y-direction decides the domain size. The cut-off radius $r_c$ represents the core region. c. Schematic representation of the homogeneous distribution of the singularity of the strain field of an edge dislocation in discrete steps. Steps 1 to 50 show the extension of singularity up to the maximum delocalization, when the dislocation can be considered to be halfway between the two adjacent equilibrium sites. For the next half, the distribution contracts again to the line singularity in steps 51 to 99. After one full cycle, the dislocation shifts by the Burgers vector. d. The existence of the potential energy barrier for a particular domain size when no shear stress has been applied.

Fig 2: Evolution of the energetics of the singularity distribution process under different applied shear stress. The profiles corresponding to $\tau_{xy}$ = 0.003 (down triangle), 0.006 (square), 0.009 (up triangle) and 0.013 (circle) are shown with the minimum of the elastic energy shifting towards the steps of more expanded distribution.

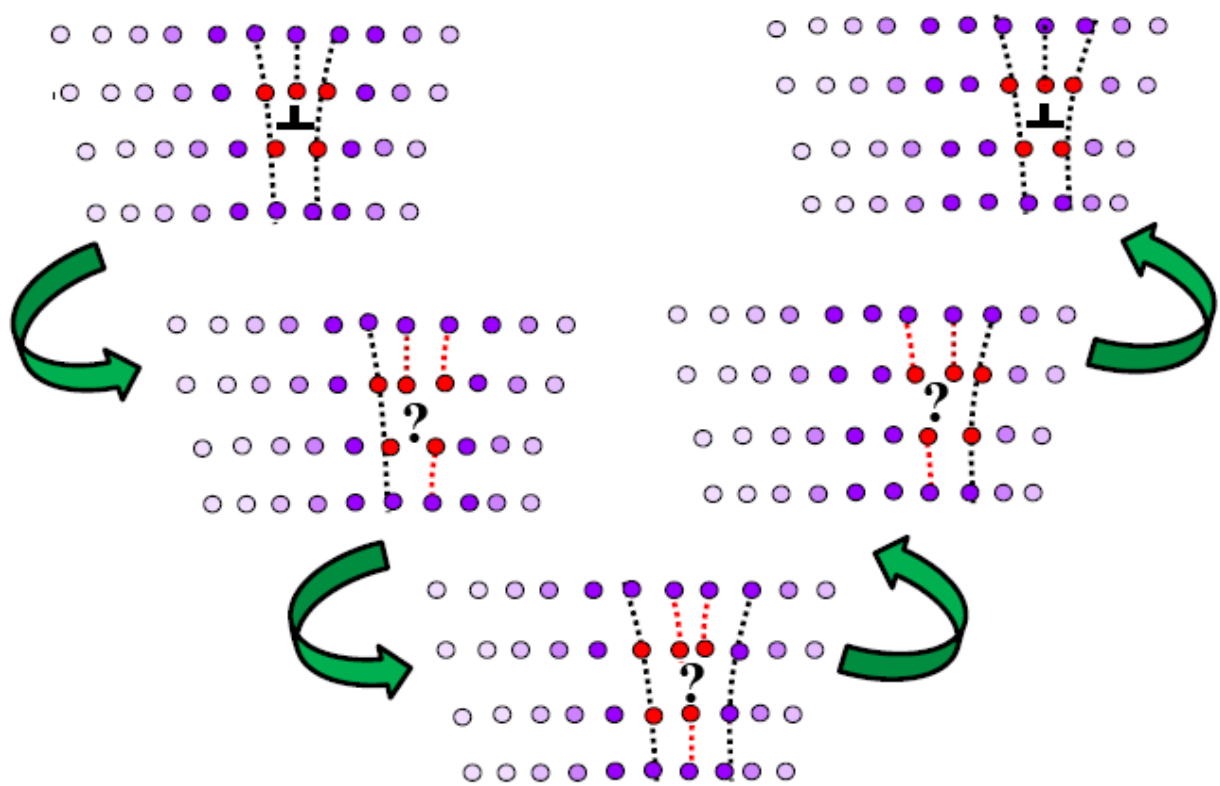

Figure 1a

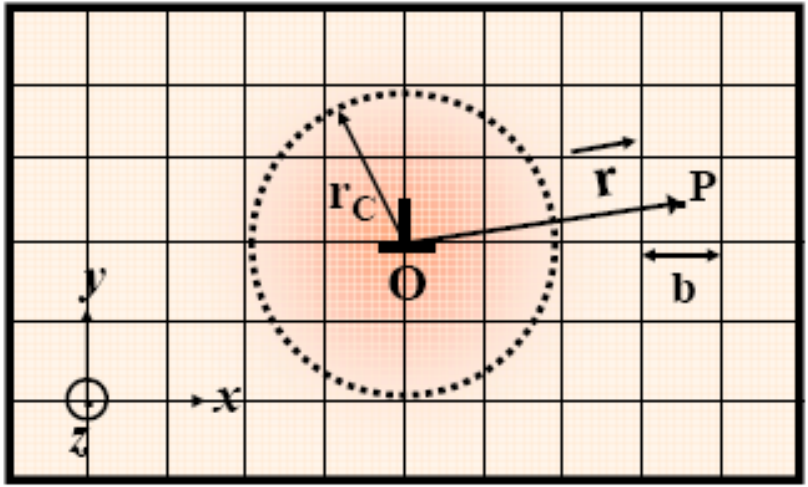

Figure 1b

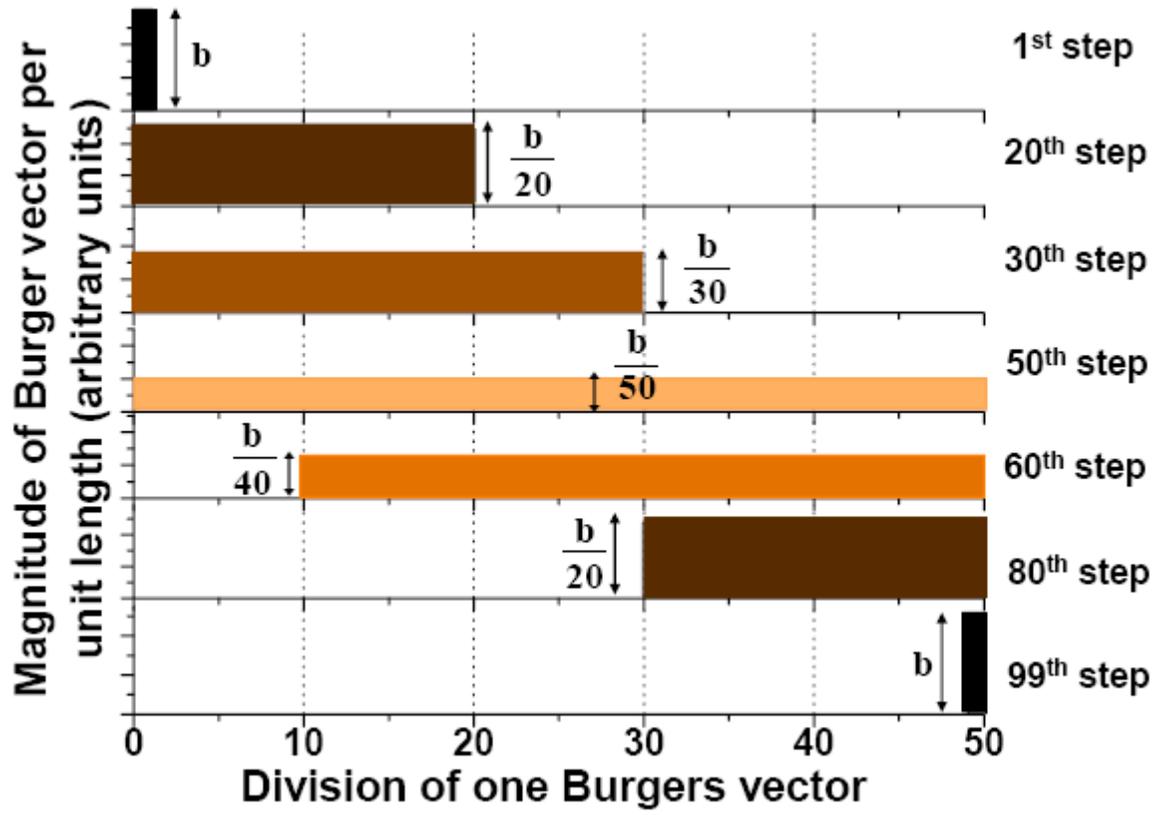

Figure 1c

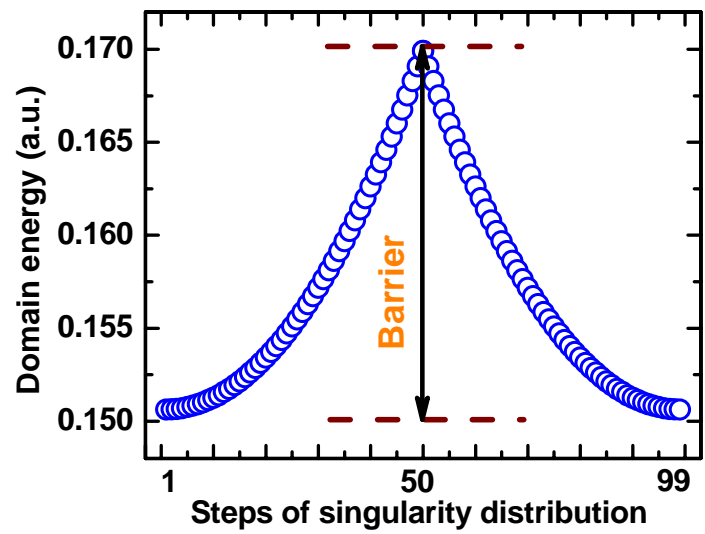

Figure 1d

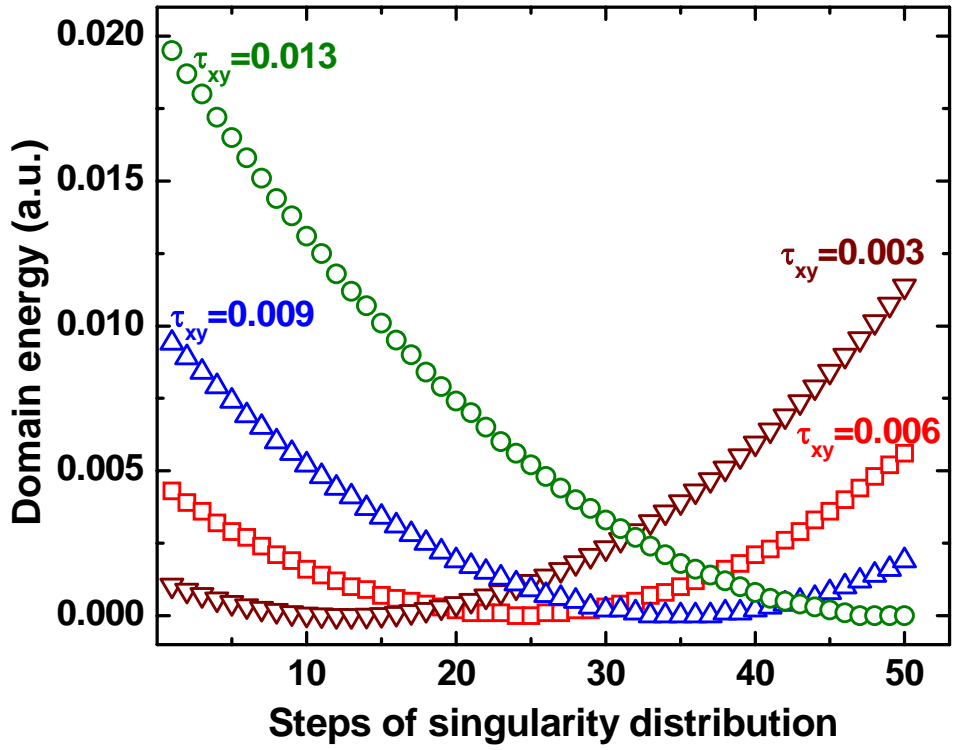

Figure 2